\documentclass[twocolumn,showpacs,aps,prl]{revtex4-1}
\usepackage{graphicx}
\usepackage{dcolumn}
\usepackage{bm}
\usepackage{epsfig}
\usepackage{amssymb}
\usepackage{metre}
\usepackage{psfrag}
\usepackage{subfigure}
\usepackage{ulem}
\usepackage{color}
\usepackage{longtable}
\usepackage[T1]{fontenc}
\usepackage{natbib}

\usepackage{array}

\begin{document}
\title{Direct and inverse pumping in flows with homogeneous and non-homogeneous swirl}

\author{A. Poth\'erat$^1$, F. Rubiconi$^1$, Y. Charles$^{2}$ and V. Dousset$^1$
}
\email[]{alban.potherat@coventry.ac.uk}
\affiliation{$^1$Coventry University, Applied Mathematics Research Centre, priory street Coventry CV1 5FB, UK}
\affiliation{$^2$Grenoble High Magnetic Field Laboratory and CRETA laboratory/CNRS-Grenoble, France}

\date{\today}

\begin{abstract}
The conditions in which meridional recirculations appear in swirling flows above a fixed wall are analysed. In the classical Bodew\"adt problem, where the swirl tends towards an asymptotic value away from the wall, the well-known 

"tea-cup effect" drives a flow away from the plate at the centre of 
the vortex. Simple dimensional arguments applied to a single vortex show 
that if the intensity of the swirl decreases away from the wall, the sense of 
the recirculation can be inverted, and that the associated flow rate scales 
with the swirl gradient. If the flow is quasi-2D, the classical 
tea-cup effect takes place. This basic theory is confirmed by numerical 
simulations of a square array of steady, electrically driven vortices. 
Experiments in the turbulent regimes of the same configuration reveal that 
 these mechanisms are active in the average flow and in 
its fluctuating part. 
These mechanisms provide an explanation for previously observed phenomena in electrolyte flows. They also put forward a possible mechanism for the generation of 
helicity in flows close to two-dimensionality, which plays 
a key role in the transition between 2D and 3D turbulence.
\end{abstract}


\maketitle

\section{Introduction}
The teacup effect is one of the mechanisms actively mixing sugar in a cup of 
tea when stirring it with a spoon. Under the effect of rotation, a centripetal 
pressure gradient builds up in the fluid to oppose the centrifugal force. In 
the boundary layer near the bottom of the cup, the flow is slow so the 
centrifugal force collapses and the same pressure gradient drives a convergent 
flow towards the centre of the cup. This feeds the meridional recirculation that is actually responsible for mixing (this configuration is that of the 
\emph{Bodew\"adt} problem). The configuration where a solid wall rotates under a 
still fluid (\emph{Ekman} problem) leads to a reversed meridional flow for the same reasons 
\cite{ekman1905_amaf,bodewadt1940_zamm}.\\
This type of mechanism is at play in a number of less anecdotal processes, 
both industrial and natural, such as the stirring of liquid metals (in the 
Bridgeman process to grow silicone crystals for example \cite{gorbunov03_jcg}) 
or the generation of cyclones, where it controls the redistribution of momentum,
 heat or chemicals in the flow. It is particularly important in experiments 
aiming to reproduce 2D turbulence or to understand the transition 
between 2D and 3D turbulence \cite{paret97_prl,sommeria86}: 
the authors of Refs.  \cite{shats10_prl,xia11_nphys} recently 
discovered that the appearance of a third velocity component played a role in 
the break-down of the inverse energy cascade of 2D turbulence, and was 
therefore central to understanding this transition. Recent numerical 
simulations \cite{biferale12_prl} were able to link the existence of the 
inverse cascade to the presence of helicity even in 3D turbulence. Since the 
 teacup effect is precisely a source of helicity linked to 
the presence of a bottom wall, it most likely plays an important role in the 
transition between 2D and 3D turbulence dynamics. The authors of Ref. \cite{akkermans08_epl} attempted to suppress it 
by inserting a "buffer" fluid layer between the container wall and the layer 
of fluid where turbulence was forced, only to discover that secondary flows 
still subsisted. Their presence was attributed to confinement itself but also 
to the non-homogeneity of the forcing, which consisted of passing an electric 
current through the fluid layer (a conductive electrolyte), placed over an 
array of magnets of alternate polarity. {The nature of these 
recirculations was not the focus of this particular paper but it can be seen from their 
results} that pumping in this 
configuration was \emph{inverse}, with the fluid diving to the centre of 
vortices in the core of the flow, whereas in homogeneously forced flows between 
two planes, \emph{direct} pumping is expected as in the teacup effect 
\cite{greenspan69}. Thus, although the teacup effect is well understood, the 
conditions in which secondary flows appear and simply which way they flow is 
not yet clear.\\ 
In this paper, we put forward a mechanism to explain how either direct or 
inverse pumping arises, depending on the homogeneity of the forcing. This
scenario is tested against numerical simulations and experiments on a flow 
of liquid metal between two parallel planes distant of $H$, subject to a 
transverse magnetic field $B \mathbf e_z$ and where the flow is driven by 
injecting electric current at one of the walls. 
This setup offers a convenient way to control the homogeneity of the forcing, 
because for strong magnetic fields, the Lorentz force diffuses momentum across 
the fluid layer \cite{sm82}. {When inertia is present, it opposes this effect 
and diffusion is only achieved over a finite length $l_z\sim l_\perp N^{1/2}$, 
where the interaction parameter $N=\sigma B^2 l_\perp/(\rho U)$ represents
the ratio of the Lorentz force to inertia ($l_\perp$, $U$, $\rho$ and $\sigma$
 are the the size of the swirling structure across 
$\mathbf B$, its azimuthal velocity, the density and electric conductivity of 
the fluid). For high magnetic fields $l_z/H>>1$ and so momentum diffusion across the channel is effective enough to make the flow quasi-2D. If, on the other hand , $l_z/H<1$ or $l_z/H\sim1$, then inertia and momentum diffusion 
are of the same order and 3D velocity variations are present \cite{pk13_jfm}.}

\section{Governing equations}
The velocity and pressure fields $\mathbf u$ and $p$ of an incompressible flow of a fluid of density $\rho$, viscosity $\nu$ are governed by the Navier-Stokes and continuity equations:
\begin{eqnarray}
(\partial_t+\mathbf u\cdot\nabla)\mathbf u+\frac1\rho\nabla p&=&\nu\nabla^2\mathbf u+\mathbf f, \label{eq:ns}\\
\nabla\cdot\mathbf u&=&0. \label{eq:cont}
\end{eqnarray}
%
To illustrate the phenomena of direct and inverse pumping, we shall rely on the 
example of electrically forced flows. The principle is to apply an externally 
generated, homogeneous magnetic field $B\mathbf e_z$ to an electrically 
conducting fluid and to inject electric current in one or several points of 
an otherwise electrically insulating wall orthogonal to $\mathbf e_z$ (here 
at $z=0$). At each such electrode, the Lorentz force creates a vortex spinning 
around  $\mathbf e_z$ that extends all the further in the core as $B$ is high 
\cite{sommeria88, pk13_jfm}. 
Assuming the magnetic Reynolds number $Rm=\mu\sigma UL$ remains small 
($U,L$ are typical velocities and length and $\mu$ is the magnetic 
permeability of vacuum.), 
then the magnetic field associated to the current 
$\mathbf J$ within the fluid is negligible compared to $B$, and the Lorentz 
force to which the flow is subjected expresses as $\mathbf f=B\rho^{-1}\mathbf J\times\mathbf e_z$ \cite{roberts67}.
 $\mathbf J$ is coupled to $\mathbf u$ through Ohm's law and charge 
conservation:
\begin{eqnarray}
\mathbf J&=&\sigma(-\nabla\phi+B\mathbf u\times\mathbf e_z),\\
\nabla\cdot\mathbf J&=&0,
\label{eq:jcons}
\end{eqnarray}
where $\phi$ is the electric potential. 
The generic geometry is that of a channel with no-slip, impermeable and 
electrically insulating walls located at $z=0$ and $z=H$. The flow is 
governed by two non-dimensional parameters: the Hartmann number 
$Ha=BH\sqrt{\sigma/(\rho\nu)}$ and the Reynolds number $Re^0=\Gamma/\nu$. 
$\Gamma=I/(2\pi\sqrt{\sigma\rho\nu})$ is the circulation that would be induced around an electrode, in the plane just outside the Hartmann boundary layer at $z=0$, by injecting a DC current of intensity $I$ through it, if no viscous dissipation was present \cite{sommeria88, pk13_jfm}. $Ha^2$ represents the ratio of Lorentz to viscous forces but $Ha$ and $Re^0$ can be thought of as non-dimensional 
measures of $B$ and $I$ respectively. Thus, $Ha$ controls the momentum diffusion along $\mathbf B$ and so 
decreasing it incurs steeper velocity gradients along $\mathbf e_z$ in the core 
of the flow, and increases the inhomogeneity of the swirl.
%
%

\section{Numerical System}
In the numerical simulations, the
fluid is confined in a rectangular box of size $L^{2}\times H$ with
$L=0.06$ m and $H=0.1$ m. 
The working fluid is GaInSn, an eutectic alloy of Gallium Indium and Tin, 
of conductivity
$\sigma=3.6\times10^{6}$ $\Omega^{-1}.$m$^{-1}$, viscosity $\nu=4\times10^{-7}$ m$^{2}.$s$^{-1}$
and density $\rho=6.4\times10^{3}$ kg.m$^{-3}$. The container is placed
in a uniform magnetic field $B\mathbf e_z$ so that besides the 
two Hartmann walls orthogonal to it, there are also four walls 
parallel to it. All
walls are impermeable, no slip and electrically insulating. The flow
is driven by injecting a DC current of alternate polarity through a square 
array of 6 $\times$
6 electrodes of diameter 1 mm, mounted flush at the bottom wall, spaced by distance
$L_{i}=0.01$ cm.\\
The code is based on the finite volumes method implemented in the 
OpenFOAM framework and solves the three-dimensional, time-dependent 
equations in a segregated way. The numerical scheme is the consistent 
and conservative 
algorithm put forward in Ref. \cite{ni07_jcp}. The
code is described and fully tested in the more complex configuration
of the flow around a 3D obstacle in Ref. \cite{dp12_jfm}. To
summarise it, the spatial discretisation is of second order, the
time-scheme is a second order implicit pressure-velocity formulation 
and the pressure-velocity coupling is solved with the PISO algorithm 
implemented as in Ref. \cite{weller98}. 
The time step is chosen
so that the maximum Courant number $C=U\Delta t/\Delta x$ is
well below 1 and the maximum of $D=\nu\Delta t/\Delta x^{2}$
remains below 10 (Courant-Friedrich-Lewy conditions). 
%
Collocated and structured meshes made of 1 917 971 and 4 227 768 
rectangular elements are respectively used for simulations at  $Ha=800$ and 
$Ha=1822$.
In order to keep the mesh orthogonal, the electrodes, which are circular 
in the experiment, are modelled by squares of size 1 mm. The mesh is regular outside of the
boundary layers and refined near the walls so as to always keep 4 points
across the Hartmann layers and 3 points across the Shercliff layers.\\ 
For each value of $Ha$, the flow is initially at rest and the lowest
current is injected. In all computations 
presented here, the flow stabilised into a steady state, 
which was used as initial condition for the simulation at
the same $Ha$, for the next value of $Re^0$ up. 
%

\section{Mechanisms of inverse and direct pumping}
\begin{figure}
\centering
\psfrag{I}{\textcolor{red}{$I$}}
\psfrag{A}{$0.17$}
\psfrag{D}{$-1/2$}
\psfrag{B}{$0$}
\psfrag{C}{}
\psfrag{E}{$0$}
\psfrag{F}{$1/2$}
\psfrag{G}{$x/L_i$}
\psfrag{X}{$x/L_i$}
\psfrag{Z}{$z/H$}
\includegraphics[width=8.5cm]{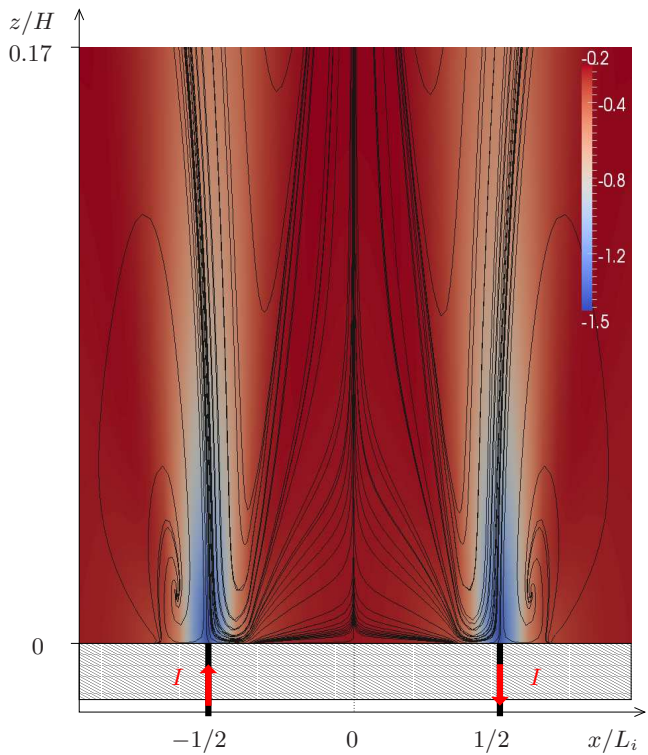}
\caption{Streamlines of $(u_x,u_z)$ in the plane $y=L_i/2$, intercepting the axis
 of rotation of two vortices and contours of pressure (colours). Pressure is 
normalised by $\rho u_M^2$ where $u_M={\rm max}\{u_y(x,z)\}$ is the maximum azimuthal velocity within any given vortex in the vessel. Downward vertical jets occur right in the region of lowest pressure at the centre of vortices at $x=\pm L_i/2$. {The fluid domain extends over $(x/L_i,y/L_i,z/H)\in[-3,3]^2\times[0,1]$, but only the region $[-1,1]\times\{1/2\}\times[0,0.17]$ is represented here.}}
\label{fig:press_stream}
\end{figure}

\begin{figure}
{\small
\hspace{-0.55cm}
\psfrag{profils}{}
\psfrag{Z}{$\tilde z$}
\psfrag{X}{$\tilde r$}
\psfrag{Utheta}{$\tilde u_y(-0.372,\tilde z)$}
\psfrag{UR}{$\tilde u_x(-0.372,\tilde z)$}
\psfrag{UZ}{$\tilde u_z(-1/2,\tilde z)$}
\psfrag{P}{$\tilde p(-1/2,\tilde z)$}
\includegraphics[width=.25\textwidth, height=8.5cm]{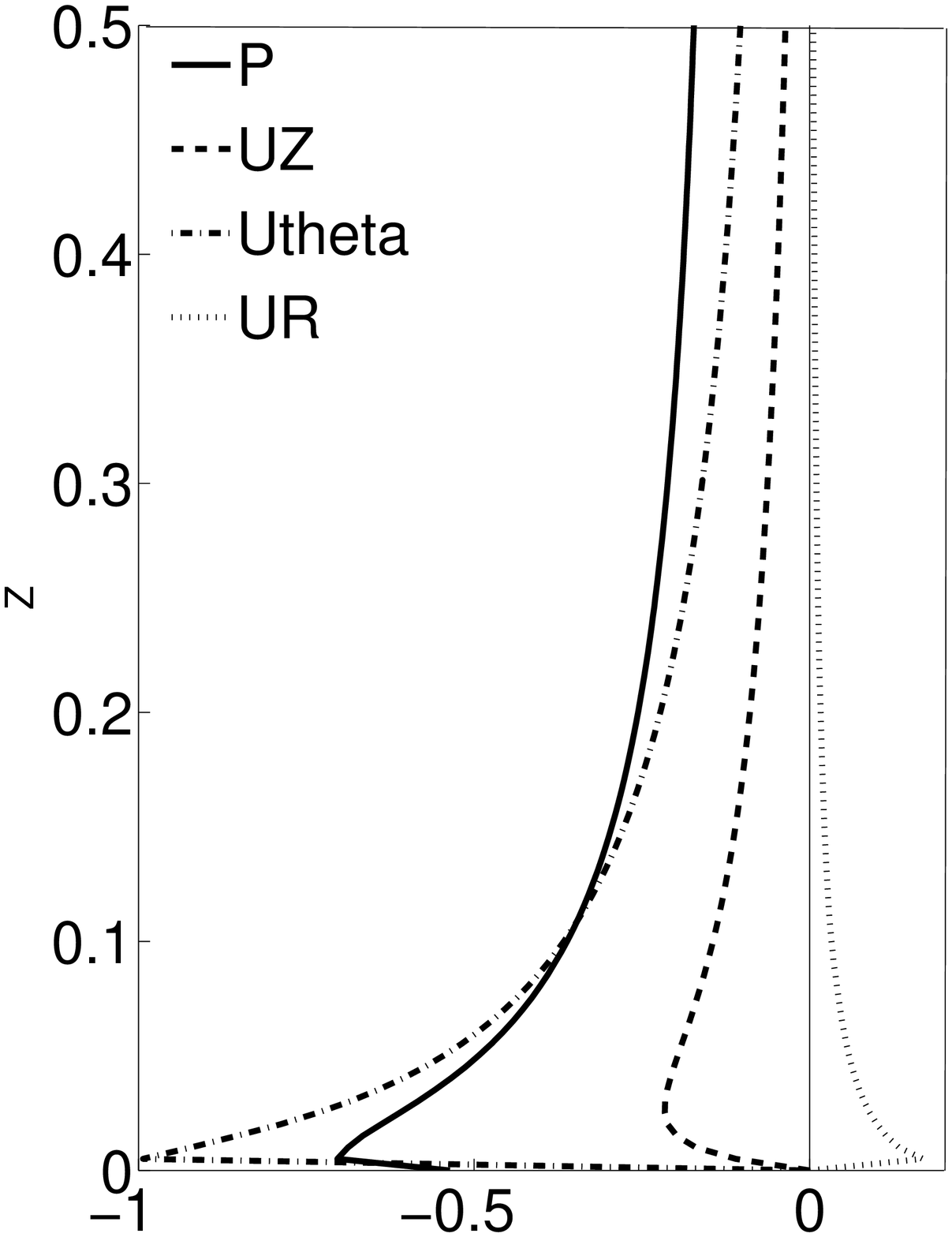}
\psfrag{nsZcomponents}{}
\psfrag{pgrad}{$\partial_{\tilde z} \tilde p$}
\psfrag{viscosity}{$\frac{\nu}{L_i u_M} \tilde\nabla^2 \tilde u_z $}
\psfrag{convAcc}{$\tilde \mathbf u\cdot\tilde\nabla \tilde u_z$}
\includegraphics[width=.25\textwidth, height=8.5cm]{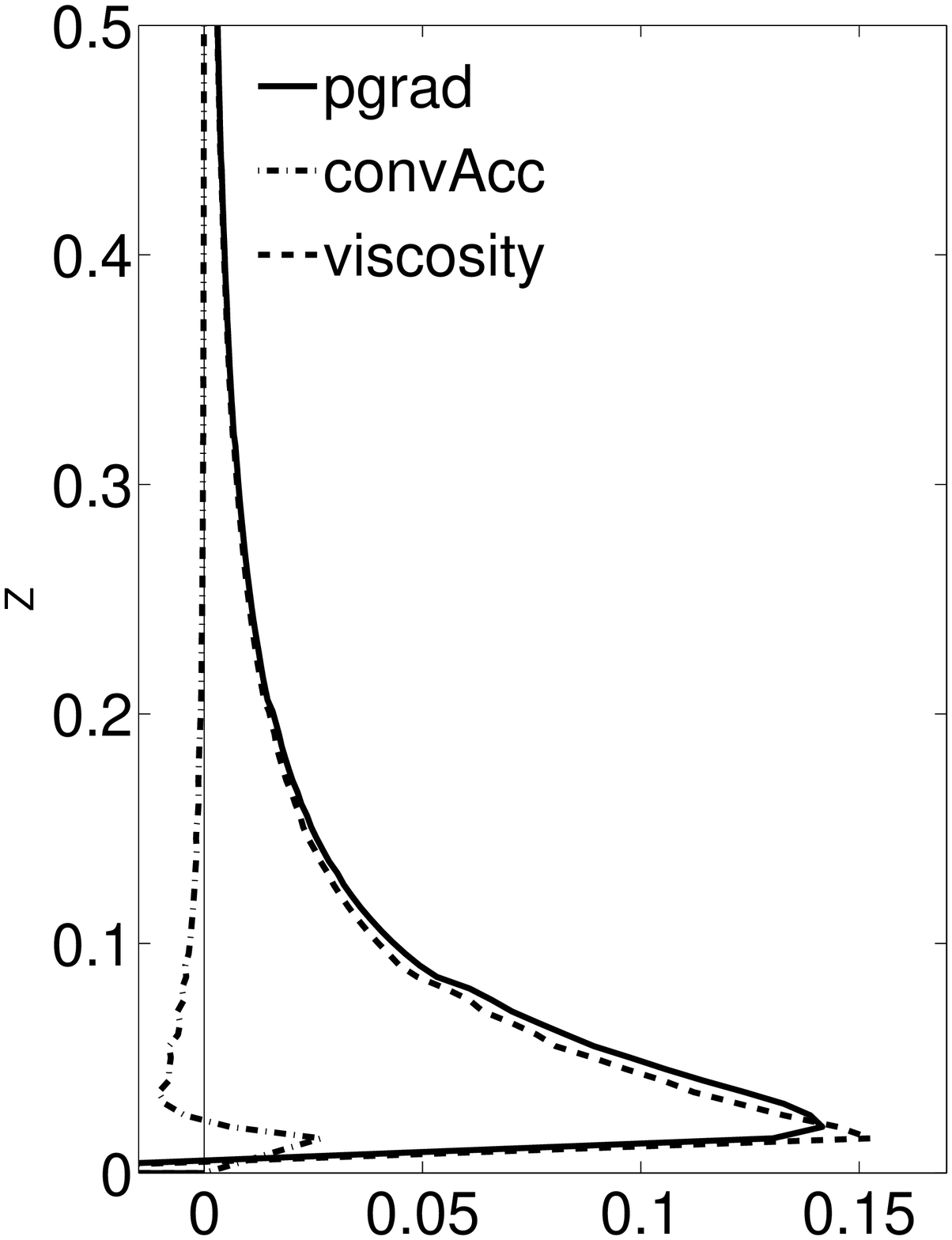}
}
\vspace{-0.5cm}
\caption{Left:  profiles of pressure and velocity in the plane $y/L_i=1/2$, 
 along the columnar vortex centred at $(x/L_i,y/L_i)=(-1/2,1/2)$. Right: variations of 
the different terms in (\ref{eq:ns})$\cdot\mathbf e_z$ along the axis of 
rotation of the same vortex. The vertical pressure gradient balances lateral 
viscous friction to a good approximation in the core. 
Velocities, pressure, horizontal and $z$ coordinates marked $\tilde q$ were 
respectively normalised by $u_M={\rm max} \{u_y(x,z)\}$, $\rho u_M^2$, $L_i$ 
and $H$. 
}
\label{fig:ns_z}
\end{figure}

We shall first illustrate the basic mechanism of inverse pumping on the example 
of a steady flow in the configuration described in the previous section. 
Figure \ref{fig:press_stream} shows the pressure contours and streamlines 
obtained at $Ha=1.822\times10^3$ and $Re^0=171$. These are typical of the flow patterns 
obtained for other parameters. The flow consists of a square lattice of 
$6\times6$ vortices rotating along $\mathbf e_z$ with alternate spin. 
A strong downward flow exists at the centre of each of them that loops back up 
in the outer part of the vortices. To illustrate the underlying mechanism, we 
shall reason on a single vortex of radius $l_\perp$, with 
 associated polar coordinates centred on it ($r^2=(x+L_i/2)^2+(y-L_i/2)^2$, and 
the radial and azimuthal velocity in this vortex respectively correspond 
to $u_x$ and $u_y$ in the plane $y=L_i/2$, on figure 
\ref{fig:press_stream}).
The pressure contours in figure \ref{fig:press_stream} and vertical profiles of 
$p$ and $u_z$  (figure \ref{fig:ns_z}, left) show that the downward jet 
coincides with a strong pressure gradient. Plots of the different terms in 
(\ref{eq:ns})$\cdot\mathbf e_z$ (figure \ref{fig:ns_z} , right)  show that the 
latter is only opposed by viscous friction: 
\begin{equation}
\partial_z p \sim -\rho\nu\frac{u_z}{l_\perp^2}.
\label{eq:nsz}
\end{equation}
The pressure drop at the centre of the vortex results from the centripetal pressure gradient that opposes the centrifugal force: 
%
\begin{equation}
 p(r=0,z)-p(r=l_\perp,z) \sim -\rho u_\theta^2(r=l_\perp,z).
\label{eq:nsr_z}
\end{equation}
Contours of $p$ on figure \ref{fig:press_stream} suggest that the pressure 
gradient outside of the vortex $\partial_z p(l_\perp,z)$ is negligible. The 
maximum $u_z^M$ of $u_z(0,z)$ is located at $z=z_0$, near the bottom wall, but 
still outside the boundary layer (see figure \ref{fig:ns_z}, left) and can be 
estimated from (\ref{eq:nsz}) and (\ref{eq:nsr_z}): 
%
\begin{equation}
\frac{u_z(0,z_0)}{u_\theta(l_\perp,z_0)}\sim Re^\nabla,
\label{eq:uz}
\end{equation}
where $Re^\nabla=2\partial_z u_\theta(l_\perp,z_0)l_\perp^2/\nu$ is a Reynolds 
number built on the vertical gradient of the swirl. 
Eq. (\ref{eq:uz}) expresses that vertical motion is driven by the inhomogeneity 
of the swirl. Mass conservation requires that the flow impacting the wall must 
turn radially:
\begin{equation}
\label{eq:ur}
u_r(z_0)\sim-\frac{l_\perp}{2h^r} u_z(0,z_0)\sim -u_\theta^o\frac{l_\perp}{2h^r} Re^\nabla, 
\end{equation}
where, $u_\theta^o=u_\theta(l_\perp,z_0)$, $h^r$ is the height over which the return flow takes place, 
typically of the same order as $l_\perp$ in all simulations.\\
Unlike inverse pumping, \emph{direct} pumping is driven in the wall boundary layer by the radial 
pressure gradient that builds up in the core to oppose the centrifugal force 
(\ref{eq:nsr_z}). If $\delta$ is 
the boundary layer thickness, (\ref{eq:ns})$\cdot\mathbf e_z$ implies that the pressure cannot vary across it and so the radial pressure gradient there is the 
same as in the core. Since $u_\theta$ becomes small in the boundary layer, so 
does the centrifugal force $\rho u_\theta^2/r$. Hence, 
(\ref{eq:ns})$\cdot\mathbf e_r$ 
there expresses that in the boundary layer, the centripetal pressure gradient
is balanced by viscous friction. If 
inverse pumping is present, however, the associated radial flow in the core, 
expressed by 
(\ref{eq:ur}), considerably 
enhances radial friction $\nu\partial_{zz}^2u_r$ in the boundary layer. Denoting quantities within the boundary layer with a 
superscript $b$, 
it comes that 
\begin{equation}
\partial_r p^b\sim -\frac{\rho\nu}{\delta^2}(u_r^b-u_r(z_0)),
\end{equation}
and from (\ref{eq:nsr_z}) and (\ref{eq:ur}),  
\begin{equation}
\frac{u_r^b}{u_\theta^o}\sim-\frac{l_\perp}{2h^r}Re^\nabla-\left(\frac{\delta}{l_\perp}\right)^2Re,
\label{eq:urh}
\end{equation}
where, unlike $Re^0$, the Reynolds number $Re=u_\theta^o l_\perp/\nu$ is 
based on $u_\theta^o$, a measured quantity.
For direct pumping to exist, the flow must be directed radially inwards within 
the boundary layer. When $\partial_z u_\theta<0$, as in the case studied here, 
this imposes the condition:
\begin{equation}
|\partial_z u_\theta|<\left(\frac{\delta}{l_\perp}\right)^2\frac{h^r}{l_\perp}\left|\frac{u_\theta^o}{l_\perp}\right|.
\label{eq:bode_pumping_condition}
\end{equation} 
In all numerical simulations, $h^r$ was typically of the order of $l_\perp$. 
Boundary layers are thin, so $l_\perp/\delta>>1$ (in electrically driven flows, 
the wall boundary layer is a Hartmann layer, with thickness $\delta/H= Ha^{-1}$,
 so $l_\perp/\delta\sim Ha$). Since the flow is quasi-2D as soon as 
$\partial_z \mathbf u=\mathcal O (\delta/l_\perp)$ outside the boundary 
layer, {rather than an exact 
criterion on $|\partial_z u_\theta|$ for the appearance of inverse pumping, 
(\ref{eq:bode_pumping_condition}) mainly expresses that inverse pumping  can only exist 
when the swirl is quasi-2D}. This result explains why direct pumping was never observed in any 
of our simulations: in all numerically accessible regimes, the 
inhomogeneity of the swirl was always sufficient to drive inverse pumping. 
Similarly, in the wall-bounded electrolytes layers studied by  Ref.
\cite{akkermans08_pf}, vortices were driven by imposing an electric current in 
the field created by permanent magnets placed underneath the layer. The swirl 
inhomogeneity resulted directly from that of the magnetic field and so  
$\partial_z u_\theta$ and $u_\theta/H$ were of the same order, making 
it impossible for (\ref{eq:bode_pumping_condition}) to be satisfied. It is 
therefore not surprising that inverse pumping was  observed in this 
case too.\\
In quasi-2D flows, by contrast, $Re^\nabla\simeq0$ and  from (\ref{eq:urh}) 
and (\ref{eq:cont}), \emph{direct} pumping drives a flow from the boundary 
layer into the core:
\begin{equation}
\frac{u_z}{u_\theta^o}\sim 2\left(\frac{\delta}{l_\perp}\right)^3 Re.
\label{eq:uz_direct}
\end{equation}
In electrically driven flows, $\delta=H/Ha$ and scaling (\ref{eq:uz_direct}) coincides with the expression derived by Ref. \cite{psm00_jfm} from matched asymptotics. 
{Ref.\cite{sommeria88} provides a clear example of \emph{direct} 
pumping in a large electrically driven vortex confined in a cylindrical vessel 
of diameter 12 cm filled with a layer of mercury of thickness $H=1.92$ cm, 
placed in axial homogeneous magnetic field of 0. 
In these conditions, the ratio of magnetic diffusion length $l_z$ to $H$ was 
much 
larger than unity (typically between $10^2$ and $6\times10^2$ ) and so the flow was indeed quasi-2D, as required by condition (\ref{eq:bode_pumping_condition}). The numerical simulations of Ref.\cite{psm05_jfm} together with the measurements of radial profiles of azimuthal velocity confirmed that direct pumping in this case scaled as (\ref{eq:uz_direct}).}\\
We shall conclude this theoretical section with two more remarks:
firstly, it should be noticed that should the swirl be inhomogeneous in 
such a way that $\partial_z |u_\theta|>0$, then direct pumping would occur, but 
$u_z$ would still scale as in (\ref{eq:uz}). {Secondly, 
(\ref{eq:bode_pumping_condition}) provides a criterion for the existence of 
\emph{direct} Bodew\"adt pumping, not a criterion for the disappearance of 
inverse pumping. In theory, one could imagine a flow that would be quasi-2D to 
a precision of $\mathcal O(\delta/H)^2$ and therefore with direct pumping, but 
with a residual three-dimensionality of higher order, which would suffice to 
drive a faint inverse pumping at the same time. More realistically, $|\partial_zu_\theta|$ may vary 
along $\mathbf e_z$, in such a way that it can satisfy (\ref{eq:bode_pumping_condition}) in the vicinity of at least one of the walls but not everywhere. The 
corresponding secondary flow could then feature a complex succession of recirculations in any direction. One such example shall be found in turbulent flows (section \ref{sec:exp}).}
\section{Inverse pumping in the square vortex array}

\begin{figure}
\psfrag{Ha1}{$Ha=8\times 10^2$}
\psfrag{Ha2}{$Ha=1.822\times 10^3$}
\psfrag{Y}{$u_z(z_0)/u_x^o$}
\psfrag{X}{$Re^\nabla$}
\psfrag{fit}{$u_z(z_0)/u_x^o=0.017 Re^\nabla$}
\includegraphics[width=8.5cm]{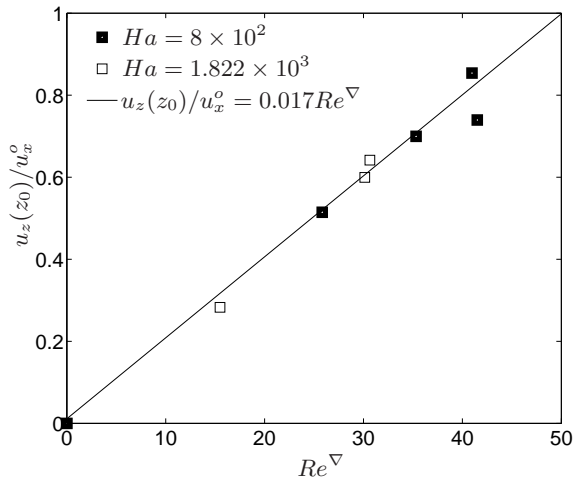}
\caption{Scaled maximum vertical velocity \emph{vs.} $Re^\nabla$, from 
numerical simulations of the $6\times6$ vortex array in steady state 
($Re^\nabla$ gives a measure of the gradient of swirl along $\mathbf e_z$). 
$Re^\nabla$ was calculated with $l_\perp=L_i/4$.}
\label{fig:uz_vs_renabla}
\end{figure}

%
%
Numerical simulations of the vortex array present specific features 
that differ from those of an isolated vortex, on which the theory is based.
Despite this difference, figure \ref{fig:uz_vs_renabla} shows that
the vertical velocity $u_z(z_0)$ within a single vortex in the array  
linearly increases with $Re^{\nabla}$. 
Scaling (\ref{eq:uz}), is thus still satisfied,
albeit with a reasonable amount of data scattering. This scattering
can be partly attributed to the plotted values being local ones, and partly to 
 interaction between vortices.  Most importantly,  
data obtained for different values of $Ha$ collapse on the same line, which 
confirms $Re^{\nabla}$ as the single relevant parameter for inverse 
pumping. This stresses that it is the gradient of swirl that determines 
inverse pumping, not how this gradient is generated, electromagnetically or in 
any other way. Furthermore, 
all simulated cases involve strongly inhomogeneous swirl, with 
$\partial_z|u_\theta| \gtrsim |u_\theta|/H$ and exhibit inverse pumping, and no 
direct pumping,  
in agreement with condition  (\ref{eq:bode_pumping_condition}) too.\\
Unlike when isolated, vortices embedded in an array undergo a strong 
influence from 
neighbouring vortices. In an infinitely extended square array, this would 
mainly translate into a loss of axisymmetry of individual vortices, with 
streamlines progressively deforming to a square shape, away from the vortex 
cores.
In our numerical simulations, by contrast, the $6\times6$ array is bounded 
by lateral walls. These incur significant friction on peripheral structures, 
which are consequently weaker, and less influent than those near the centre. 
As a consequence, 
the upper part of the vortex axes is slightly 
diverted away from the centre, instead of being straight.  
Nevertheless, as vortices are driven by injecting current from
the bottom wall, they remain attached to the electrodes in this region, 
and the effect of the imbalance between centre and peripheral vortices becomes more 
visible away from the bottom wall (This can be noticed in figure \ref{fig:press_stream}). 
A second property of vortex arrays is that radial flows induced in 
neighbouring vortices by inverse pumping collide to form a strong 
return flow that spirals up in the region between vortices (This pattern is 
reflected in the vertical jet at $x=0$ in the streamlines 
of $(u_x,u_z)$ in figure \ref{fig:press_stream}).
When $Re^0$ is increased, these phenomena become more pronounced, and at $Re^0=171$ (for $Ha=800$) and $Re^0=512$ (for $Ha=1822$), 
recirculations merge in the upper part of the vessel, to the point where they 
 cannot be distinguished from each other anymore.

%
\section{Direct and inverse pumping in turbulent flows}
\label{sec:exp}
We shall now illustrate the occurrence of inverse pumping in more complex flows, 
such as turbulent flows. Since the corresponding regimes lie beyond the reach 
of numerical simulations, these shall be analysed experimentally.
The experimental setup closely matches the configuration of the numerical 
simulations with one difference: the dimensions of the vessel across the magnetic field are $L\times L=0.1$m$\times0.1$m instead of $L\times L=0.06$m$\times0.06$m. The distance between electrodes where the current is injected can be set to $L_i=0.01$m (as in the numerical simulations) or $L_i=0.03$m. A full 
description of the rig can be found in \cite{kp10_prl,pk13_jfm}. Since the 
influence of the walls becomes more pronounced when $L_i/L$ increases, 
the numerical simulations present an intermediate 
case between the two injections scales available in the experiment. For the 
purpose of this work, the rig was equipped with ultrasound sensor-transducers 
fitted flush in the top wall and in one of the lateral walls, connected either 
to a DOP1000 or a DOP3010 ultrasound velocimeter, manufactured by SIGNAL 
PROCESSING. Ultrasound velocimetry is the method of choice to obtain 
instantaneous velocity profiles in opaque fluids and is now well developed for 
liquid metal flows \cite{brito01_ef,brito11_pre}. These probes provide 
instantaneous profiles of $u_z$ along $\mathbf e_z$ 
at $(x,y)=(-1.5 {\rm cm}, 1.5{\rm cm})$ (probe V1, aligned with a current 
injection electrode), as well as 5 horizontal profiles of $u_x$ along 
$\mathbf e_x$ at 
$z\in\{0.12H,\quad 0.31H,\quad 0.5H,\quad 0.69H,\quad 0.88H\}$ (probes H1 to 
H5), also halfway between two electrodes. This gives us access to{ the variations of horizontal velocity}. The signals 
were reliable up to 
resolutions of 15Hz and 2mm. In subsequent experiments, a constant current is 
injected in a fluid initially at rest and all results presented thereafter are 
obtained when the flow has reached a statistically steady state. The 
flow is in a turbulent state, where fluctuations exceed the 
intensity of the average flow.\\
\begin{figure}
\psfrag{zh}{$z/H$}
\includegraphics[width=8.5cm]{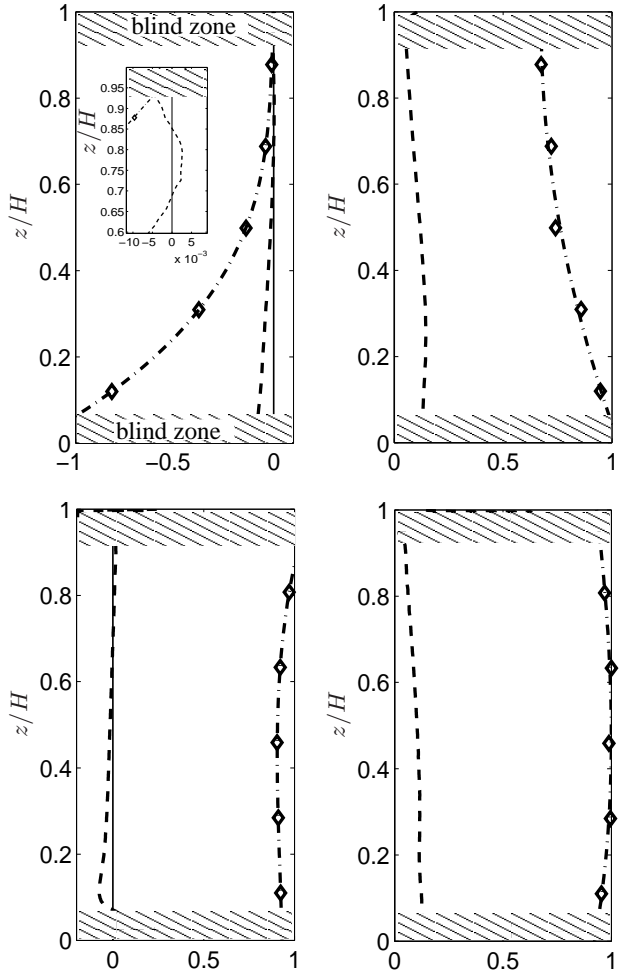}
%
\caption{Profiles of time-averaged horizontal and vertical velocities (left) 
and RMS  of velocity fluctuations (right)  obtained 
by ultrasound velocimetry. 
Average velocities (\emph{resp.} fluctuations) are normalised by ${\rm max}\{\langle u_x(x,z)\rangle\}$, (\emph{resp.} ${\rm max}\{\langle u_x(x,z)^{\prime 2}\rangle^{1/2} \}$). Dash: vertical velocity, diamonds: value of $u_x$ (average and fluctuations) measured at $(x/H,y/H,z/H)=(0,0,z/H)$, dash-dot: order 3 polynomial fit for $u_x$, 
top: $Ha=1.0932\times10^4$, $Re^0=2.488\times10^4$, $L_i/H=0.1$, 
bottom: $Ha=7.288\times10^3$, $Re^0=3.11\times10^3$, $L_i/H=0.3$. {The insert in the upper left graph represents a magnification of the profiles in the vicinity of the upper blind zone, stressing the existence of a small zone where $u_z>0$.}}
\label{fig:profiles_avg} 
\end{figure}
Figure \ref{fig:profiles_avg} shows the time-averaged profiles of 
$u_z(-1.5 {\rm cm},1.5 {\rm cm},z)$ obtained from probe V1, and $u_x(0,0,z)$ 
from probes H1 to H5, 
as well as the RMS of 
the fluctuations of these quantities (respectively denoted as $\langle\cdot\rangle$ and $\langle\cdot^{\prime2}\rangle^{1/2}$).
For $L_i/H=0.1$, $Ha=1.0932\times10^4$ and $Re^0=2.488\times10^4$, the average 
and the fluctuations of the swirl are clearly  3D, as $|u_x(z)|$ noticeably 
decreases away from the bottom wall. Inverse pumping is present in the lower 
half of the vessel, in agreement with 
the prediction of (\ref{eq:bode_pumping_condition}). The general shape of 
 profile $\langle u_z(z)\rangle$ qualitatively follows (\ref{eq:uz}), in the 
sense that $\langle u_z(z)\rangle$ is more intense where the gradient of swirl 
is stronger.
Furthermore, faint 
\emph{direct} pumping can be noticed in the 
vicinity of the upper plate (with the flow directed \emph{away from} the plate, {see insert in figure \ref{fig:profiles_avg}}). 
This local effect takes place in a region where both the swirl and its vertical 
gradient are  weak. Over a short range of values of $z$, between the regions of 
direct and inverse pumping, $u_z(z)$ becomes slightly positive, suggesting that 
a counter-rotating "connecting" recirculation is present between the 
co-rotating recirculations associated to direct and inverse pumping. The 
profile of fluctuations does not provide the direction of the vertical 
fluctuating flow, but  still shows it is strong. Its intensity is stronger in 
regions of stronger vertical gradients of 
$\langle u_x(0,0,z)^{\prime 2}\rangle^{1/2}$, which suggests that it is also 
driven by the inverse pumping mechanism.\\
For $L_i/H=0.3$, $Ha=7.288\times10^3$ and $Re^0=3.11\times10^3$, larger, slower 
vortices are more sensitive to momentum diffusion by the Lorentz force and the 
flow is correspondingly closer to quasi two-dimensionality. 
{Plots of horizontal velocity in figure \ref{fig:profiles_avg} 
show that} the average flow nevertheless still displays 
the trace of the forcing, with velocities that are highest near the bottom wall,
 high near the top wall (because of the strong electric current present in the 
top Hartmann layer) and weaker in the middle, where inertial effects take away  
energy of the main flow (The mechanisms governing three-dimensionality in wall 
bounded MHD flows are analysed in detail in \cite{pk13_jfm}). Remarkably, this 
small three-dimensionality {in the horizontal velocity profiles} is 
sufficient to drive inverse pumping in the 
vicinity of both top an bottom walls and supersede direct Bodew\"adt pumping, 
in line with our theoretical prediction that direct pumping can only occur 
in quasi-2D flows.
The profile of $\langle u_x(0,0,z)^{\prime 2}\rangle^{1/2}$ 
is practically quadratic and symmetric with respect to $z/H=1/2$. The
maximum intensity of the fluctuations {of horizontal velocity} 
corresponds to the minimum of those of the average, from which 
they draw energy (near $z/H=1/2$). This phenomenon, called "Barrel effect", was 
first predicted theoretically \cite{psm00_jfm}, then found numerically \cite{muck00_jfm} and was recently shown to act as a general mechanism of appearance 
of three-dimensionality in wall bounded flows \cite{p12_epl}. 
These measurements constitute the first experimental observation of this effect. The vertical velocity is strongly fluctuating too, but the low value of 
the gradients of $\langle u_x(0,0,z)^{\prime 2}\rangle^{1/2}$ does not allow 
us to conclude as to whether the fluctuations correspond to direct or inverse 
pumping in this case.
\section{Conclusion}
Theory, numerical simulations and experiments concur to show that direct 
recirculations can be inverted (or reinforced)  when velocity gradients appear 
in the third direction. 
This 
phenomenon does not only take place in steady vortices but also in fluctuating 
structures, which raises the question of its relevance to fluctuations at the 
different scales of turbulent flows. The appearance or the inversion 
of secondary flows in turbulent structures indeed provide mechanisms to 
create helicity or to reverse its sign. Since the sense of the energy cascade 
is tightly linked to signed helicity \cite{biferale12_prl}, they could play 
an important role in the transition between direct and inverse energy cascades. 
Nevertheless, our early experimental 
observations of turbulent flows show that inverse pumping combines with other 
effects to drive complex flow patterns with an helicity that can change sign 
along the transverse direction in a far from straightforward way.\\

%
The authors are indebted to The CRETA and LnCMI CNRS laboratories in Grenoble, 
and in 
particular, to Dr. Andr\'e Sulpice and Dr. Fran\c cois Debray for hosting and 
supporting the experiment, to Dr. Antoine Alemany (SIMAP-Madylam/CNRS) and 
Dr. Henri-Claude Nataf (ISTERRE/CNRS) for making their ultrasound velocimeters 
available to them over extended periods of time.
\bibliographystyle{nature}
\setlength\bibsep{0.pt}
\bibliography{fullbiblio}

\end{document}